\documentclass[%
preprint,
showpacs,
amsmath,amssymb,
aps,
]{revtex4-1}

\usepackage[dvipdfmx]{graphicx}
\usepackage{bm}
\usepackage{setspace}

\begin{document}

\preprint{}

\title{
Correlation function for generalized P\'{o}lya urns:
\\
 Finite-size scaling analysis
}

\author{Shintaro Mori}
 \email{mori@sci.kitasato-u.ac.jp}
 \affiliation{Department of Physics, Kitasato University \\
Kitasato 1-15-1, Sagamihara,  Kanagawa 252-0373, JAPAN
}

\author{Masato Hisakado}
\affiliation{
Financial Services Agency
\\
Kasumigaseki 3-2-1, Chiyoda-ku, Tokyo 100-8967, Japan
}%

\date{\today}

\begin{abstract}
We describe a universality class for the transitions of 
a generalized P\'{o}lya urn by studying  
the asymptotic behavior of the normalized 
correlation function $C(t)$ using finite-size scaling analysis.  
$X(1),X(2),\cdots$ are the 
successive additions of a red (blue) ball ($X(t)=1\,(0)$) at stage 
$t$ and $C(t)\equiv \mbox{Cov}(X(1),X(t+1))/\mbox{Var}(X(1))$.
Furthermore, $z(t)=\sum_{s=1}^{t}X(s)/t$ represents the successive proportions of 
red balls in an urn to which, at the $t+1$-th stage, a red ball is 
added ($X(t+1)=1$) with probability $q(z(t))=(\tanh [J(2z(t)-1)+h]+1)/2,
J\ge 0$, and a blue ball is 
added ($X(t+1)=0$) with probability $1-q(z(t))$.  
A boundary $(J_{c}(h),h)$ exists in the $(J,h)$ plane
between a region with one stable fixed point and 
another region with two stable fixed points for $q(z)$.
$C(t) \sim c+c'\cdot t^{l-1}$ with $c=0\,(>0)$ for $J<J_{c}\,(J>J_{c})$, 
and $l$ is the (larger) value of the slope(s) of $q(z)$ at the 
stable fixed point(s). On the boundary $J=J_{c}(h)$, 
$C(t)\simeq c+c'\cdot (\ln t)^{-\alpha'}$ and 
$c=0\,(c>0), \alpha'=1/2\,(1)$ for $h=0\,(h\neq 0)$.
The system shows a continuous phase transition for $h=0$ and
$C(t)$ behaves as $C(t)\simeq (\ln t)^{-\alpha'}g((1-l)\ln t)$ with a
universal function $g(x)$ and a length scale $1/(1-l)$ with respect to 
$\ln t$. 
$\beta=\nu_{||}\cdot \alpha'$ holds with
$\beta=1/2$ and $\nu_{||}=1$. 
\end{abstract}

\pacs{
05.70.Fh,89.65.Gh
}
\maketitle


\section{\label{sec:intro}Introduction}
The contagion process is one of the most studied topics 
in statistical physics and has 
attracted the attention of many researchers from various disciplines
\cite{Liu:2014,Gra:2014,Gon:2011,Cas:2009,Cur:2006,Wat:2002,Con:2000,Lux:1995}.
 P\'{o}lya urn is one of the simplest models for this process 
\cite{Pol:1931,His:2006,Hui:2008}.
 In this model, an urn consists of $t$ balls, where the proportion of red balls is $z(t)\in (0,1)$ 
 and the rest of the balls are blue. 
 The probability of a new red ball being 
 added to the urn  
 is $z(t)$, while it is $1-z(t)$ for a new blue ball; 
 the proportion of red balls then becomes $z(t+1)$. 
 This procedure is iterative, which produces a 
 sequence of proportions $z(t_{0}),z(t_{0}+1),z(t_{0}+2),\cdots$, where 
 the urn contained $t_{0}\cdot z(t_{0})$ red balls at $t=t_{0}$.   
 The limit value $\lim_{t\to \infty}z(t)$ obeys a beta distribution 
 with shape parameters $\alpha=t_{0}\cdot z(t_{0})$ and 
 $\beta=t_{0}\cdot z(t_{0})$. 

 As the P\'{o}lya urn process is very simple and 
 there are many reinforcement 
 phenomena in nature and the social environment, many variants of the process 
 have been proposed, referred to as generalized P\'{o}lya urn processes\cite{Pem:2007}. In the nonlinear generalizations of this model, 
 a continuous function $q:[0,1]\to [0,1]$ determines the 
 probability $q(z(t)) $ of a red ball being added at stage $t+1$. 
 This nonlinear version is referred to as a nonlinear 
 P\'{o}lya process\cite{Hil:1980,Pem:2007,Pem:1991}.
 In contrast to the original linear model, the nonlinear model 
 can have many isolated stable states.
 The fixed point $z_{*}$ of $q(z)$, where $q(z_{*})=z_{*}$, 
 is (un)stable if $(z-z_{*})(q(z)-z)$  is 
 negative (positive) for all $z$ in the 
 vicinity of $z_{*}$ \cite{Hil:1980}.
 $z_{*}$ is referred to as downcrossing (upcrossing) 
 as the graph $y=f(z)$ crosses the curve $y=z$ in the  
 downwards (upwards) direction.  
 The slope of $q(z)$ at $z_{*}$ is smaller (larger) than one when 
 $z_{*}$ is downcrossing (upcrossing).
 When  $q(z)$ touches the diagonal $q=z$ in the $(z,q)$ plane 
 at $z_{t}$, a point that is referred to as the touchpoint, 
 the stability of $z_{t}$ depends on the difference between the 
 slope of $q(z)$ and the diagonal $z$ in the left neighborhood 
 of $z_{t}$\cite{Pem:1991}. 
 If it is less (more) than $1/2$, $z_{t}$ is (un)stable.
 The multiplicity of the stable states provides a convenient 
 picture that explains the
 lock-in phenomena in the technology and 
 product adoption processes \cite{Art:1989}.
 Suppose two selectively neutral technologies enter the 
 market at the same time.
 Because economies of scale play the role of an externality that 
 persuade a new consumer to buy the dominant technology, 
determining which technology to buy depends on the proportion of each technology
 possessed by previous consumers. If the dependence is described 
 by the non-linear function $q(z)$, the technology adoption process 
 is described by a non-linear P\'{o}lya urn.
 An S-shaped $q(z)$ function with two stable fixed points 
 suggests the random monopoly formed when one technology 
 dominating over the other depends on chance fluctuations at the start
 \cite{Art:1987}.

 Information cascade provides a good experimental setup for 
 verifying theoretical predictions 
\cite{Bik:1992,And:1997,Mor:2012,Mor:2013}. 
 Here, participants sequentially answer questions with 
 two possible choices (two-choice questions). 
 In addition to their own information or 
 knowledge, they refer to social information about
 the number of previous subjects that chose
 each option. In an experiment where 
 subjects answer general knowledge two-choice questions, 
 it is possible to change the number of stable states 
 by controlling the difficulty 
 of the question \cite{Mor:2012,Mor:2013}. 
 A subject that knows the answer 
 to a question chooses the correct answer with a probability of 1.
 A subject who does not know the answer 
 tends to choose the majority choice. 
 By changing the difficulty of the question, an experimenter
 can control the ratio $p$ of the latter, no-knowledge, subject.
 We denote the 
 probability that the no-knowledge subject choose the correct 
 answer as $q_{h}(z)$, when the ratio of correct choices among 
 previous subjects is $z$.
 The probability that a subject choose the correct answer
 is then $q(z)=(1-p)\cdot 1+p\cdot q_{h}(z)$. 
 The sequential voting process in the experiment 
 is described by a non-linear P\'{o}lya urn. 
 It was shown that $q(z)$ has one (two) 
 stable fixed point(s) for $p<p_{c}(>p_{c})$. 

 If there is only one stable fixed point, $z(t)$ converges to it.
 In the case of multiple stable fixed points,
 the stable fixed point to which
 $z$ converges will be random\cite{Hil:1980}. 
 By controlling the model parameters, the number of stable fixed points 
 can be changed, which induces a non-equilibrium transition. 
 In an exactly solvable case, where $q(z)$ is a combination of 
 the constant $q_{*}\in (1/2,1]$ and the Heaviside 
 function $\theta(z-1/2)$, i.e., $q(z)=(1-p)\cdot q_{*}
 +p\cdot\theta(z-1/2)$ with a 
 correlation control parameter $p \in [0,1]$, $q(z)$ touches the diagonal at 
 $z_{t}=1/2$ for $p=p_{c}=1-1/2q_{*}$. For $p<p_{c}$, there is a unique
 stable state at $z_{+}=(1-p)q_{*}+p$. For $p>p_{c}$, there are two 
 stable states at $z_{\pm}=(1-p)q\pm p$.
 Because the slope of $q(z)$ in the left neighborhood of 
 $z_{t}$ is 0, the touchpoint  at $z_{t}$ is unstable, 
 and $z(t)$ converges to the stable fixed point 
 at $z_{+}$ for $p=p_{c}$ \cite{His:2011}.
  
 The probability of convergence 
 to a stable fixed point depends 
 strongly on the color of the first ball 
 when there are multiple stable states.
 If the color is red and $z(1)=1$ (blue and $z(1)=0$), the probability of 
 convergence to a larger stable fixed point 
 becomes higher (lower). The difference in the probabilities 
 is given by the limit value $c$ of the 
 normalized correlation function $C(t)$ between the color 
 of the first ball and that of the $t+1$-th ball. 
 Furthermore, $c$ plays the role of the order parameter for the phase 
 transition. In the aforementioned exactly solvable model, which we 
 refer to as the "digital" model, $C(t)$ at $p=p_{c}$ shows a power law 
 dependence on $t$ as $C(t)\propto t^{-\alpha}$ with $\alpha=1/2$. 
 The order parameter behaves as 
 $c\propto (p-p_{c})^{\beta}$ with $\beta=1$.
 In addition, $C(t)$ obeys the scaling form 
 $C(t) \propto t^{-\alpha}g(t/\xi)$ 
 near $p_{c}$ with a universal 
 function $g$ and correlation length $\xi$. 
 $\xi$ diverges as $\xi\propto |p-p_{c}|^{-\nu_{||}}$ with $\nu_{||}=2$
 \cite{Mor:2015}.
 The scaling relation $\beta=\nu_{||}\cdot \alpha$  holds as in 
 the absorbing states phase transition \cite{Hin:2000,Odo:2004}.

 In this work, we use finite-size scaling (FSS) 
 analysis in order to study the asymptotic behavior of 
 the correlation function for generalized P\'{o}lya urns.
 We adopt a logistic-type model $q(z)=(\tanh [J(2z-1)+h]+1)/2$ 
 with two parameters $J$ and $h$. 
 Here, $J$ is the parameter of the strength of the correlation, 
 and $h$ is the parameter of the asymmetry.
 The motivation for adopting this model was derived 
 from experimental findings \cite{Mor:2012,Mor:2013}.
 With this choice, there is a threshold value $J_{c}(h)$, 
 and at $J=J_{c}(h)$, $q(z)$ becomes tangential to
 the diagonal at $z_{t}$. From the above discussion, the touchpoint at
 $z_{t}$ is stable, which 
 differs from the digital model. 
 There are two stable states at $z_{t}$ and $z_{+}$ for $h\neq 0$.  
 If the order parameter $c$ takes a positive value 
 at $J=J_{c}(h)$, the phase transition becomes 
 discontinuous. For $h=0$,  
 the touchpoint at $z_{t}=1/2$ is the unique stable state. 
 $c$ should be equal to 
 zero and  the phase transition becomes continuous. 
 We also clarify the universality class of the continuous transition.     

 The remainder of the paper is organized as follows.
 Section \ref{sec:model} introduces the model.
 In section \ref{sec:cor}, we discuss the  asymptotic behavior 
 of $C(t)$ for $J\neq J_{c}(h)$ using previous results, and 
 propose the asymptotic form 
 $C(t)\simeq c+c'(\ln t)^{-\alpha'}$  for $J=J_{c}(h)$.  
 In section \ref{sec:fss}, 
 we study the FSS relations of the system. 
 Section \ref{sec:fss_Jc} is devoted to the FSS study of $C(t)$
 for $J=J_{c}(h)$. 
 We show that $c>0\,\,\, (c=0)$ and $\alpha'=1\,\,\,
 (\alpha'=1/2)$ for $h>0\,\,\,(h=0)$. The system shows  a continuous 
 phase transition for $h=0$.
 In section \ref{sec:UC}, we study the universality class of 
 the continuous transition. 
 We show that $C(t) \propto (\ln t)^{-\alpha'}g(\ln t/\xi)$ with a 
 universal function $g(x)$ and a length scale $\xi$.
 We define the critical exponents $\beta$ and $\nu_{||}$ as
 $c\propto (J-J_{c})^{\beta}$ and $\xi\propto |J-J_{c}|^{-\nu_{||}}$,
 respectively.
 Using the scaling relation $\beta=\nu_{||}\cdot \alpha'$ 
 with $\nu_{||}=1$, we obtain $\beta=1/2$.
 Section \ref{sec:con} provides our summary and further comments.
 In Appendix A, we derive the explicit form of $g(x)$ for 
 the digital model.

\section{\label{sec:model}Model}
We define the stochastic process 
$X(t)\in \{0,1\},t\in \{1,2,\cdots,T\}$, 
where the probability 
that $X(t)$ takes value of 1 is given by a function $q(z)$
of the proportion  $z(t-1)$ of the variables $X(1),\cdots,X(t-1)$
that are equal to 1. 
\begin{eqnarray}
q(z)&\equiv& \mbox{Pr}(X(t)=1|z(t-1)=z)=\frac{1}{2}
\left(\tanh \left[J(2 \cdot z-1)+h\right]+1\right),  \nonumber \\
z(t)&=&\frac{1}{t}\sum_{s=1}^{t}X(s)\,\,\,\, \mbox{for} \,\,\,\, t>0 , \,\, \mbox{and}\,\, 
z(0)=\frac{1}{2} 
\label{eq:model}.
\end{eqnarray}
The choice of $q(z)$ is arbitrary, and we adopt the above form, which is 
familiar in the field of  physics 
\cite{Sta:1971,His:2015}. 
The fixed point of $q(z)$ is a solution to 
$q(z)=z$. Using the mapping $m=2z-1$, we obtain the self-consistent equation 
$m=\tanh(J\cdot m +h)$ for the magnetization $m$ in the mean-field Ising model.
Here, we consider only the case for which $J\ge 0$. 
Because of the symmetry under 
$(X,h) \leftrightarrow (1-X,-h)$, we also assume that $h\ge 0$.

\begin{figure}[htbp]
\begin{center}
\includegraphics[width=8cm]{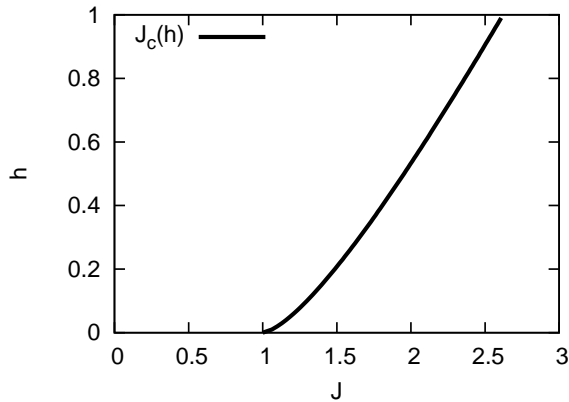}
\end{center}
\caption{Plot of $J_{c}(h)$ in $(J,h)$.}
\label{fig:Jc}
\end{figure}	

\begin{figure*}[htbp]
\begin{tabular}{cc}
\includegraphics[width=8cm]{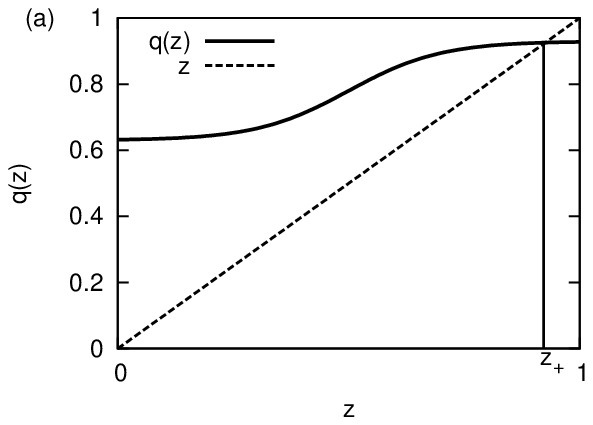} &
\includegraphics[width=8cm]{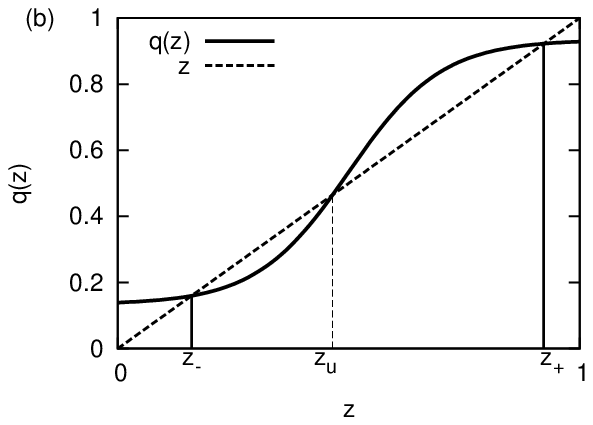} \\
\end{tabular}
\caption{Plot of $q(z)$ vs. $z$ for $h>0$.
The intersection between $y=q(z)$ and $y=z$ is the fixed point of 
$q(z)$. (a) $J<J_{c}(h)$, with one fixed point at $z=z_{+}$.
(b) $J>J_{c}(h)$, with  three fixed points at 
$z\in \{z_{-},z_{u},z_{+}\}$.}
\label{fig:ONE-TWO}
\end{figure*}	

\begin{figure*}[htbp]
\begin{tabular}{cc}
\includegraphics[width=8cm]{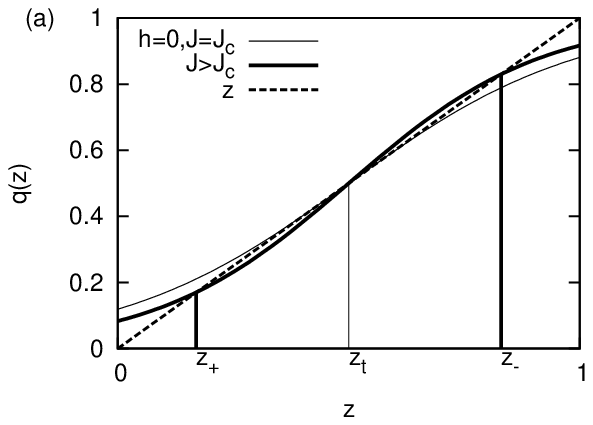} &
\includegraphics[width=8cm]{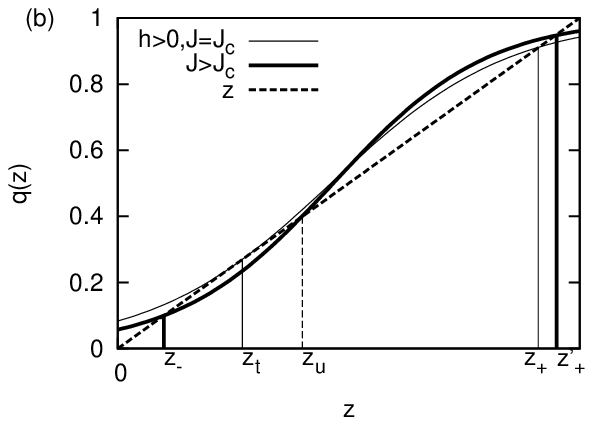}
\end{tabular}
\caption{Plot of $q(z)$ vs. $z$ for $J=J_{c}(h)$ (thin solid line) 
 and $J>J_{c}(h)$ (thick solid line)
 for (a) $h=0$ and (b) $h>0$.
}
\label{fig:Bifurcation}
\end{figure*}

The number of fixed points for $q(z)$ depends on $(J,h)$.
There is a threshold value $J=J_{c}(h)$ as a function of  $h$ 
(Fig. \ref{fig:Jc}). 
For $J<J_{c}(h)$, there is only one fixed point 
at $z=z_{+}$ (Fig.\ref{fig:ONE-TWO}(a)). 
With increasing $J$, $q(z)$ becomes tangential
to the diagonal $z$ at $z_{t}$ for $J=J_{c}(h)$. 
For $h>0$, $z_{t}\neq z_{+}$, and both $z_{t}$ and $z_{+}$ 
are stable (Fig.\ref{fig:Bifurcation}(b)). 
For $h=0$, $z_{t}=z_{+}$, and it is also stable
(Fig.\ref{fig:Bifurcation}(a)).
In both cases, the slope of the curve at $z_{t}$ is equal to one.
For $J>J_{c}(h)$, there are three fixed points, and we denote them 
as $z_{-}<z_{u}<z_{+}$; $z_{\pm}$ is stable, and $z_{u}$ is unstable 
(Fig.\ref{fig:ONE-TWO}(b)).
We denote the value $q(z_{\pm}),q(z_{t})$ as $q_{\pm},q_{t}$
and the slope of $q(z)$ at $z_{\pm},z_{t}$ as
$l_{\pm}\equiv q'(z_{\pm}),l_{t}=q'(z_{t})=1$. 
As $z_{\pm}$ is stable and downcrossing, $l_{\pm}<1$.

We note a crucial difference between $h=0$ and $h>0$.
For $h=0$, the touchpoint at $z_{t}=1/2$ for $J=J_{c}(0)=1$
coincides with the stable fixed point at $z_{+}=1/2$ for $J<J_{c}(0)$.
It splits into the two stable fixed points at $z=z_{\pm}$ for $J>J_{c}(0)$.
 (Fig.\ref{fig:Bifurcation}(a)).
$z_{\pm}$ continuously moves away from $z_{t}$ and 
$z_{+}-z_{-} \propto (J-J_{c})^{1/2}$  for $|J-J_{c}|<<1$
as in the case of the mean field Ising model.
 On the other hand, for $h>0$, the touchpoint $z_{t}$
 appears at a different position from $z_{+}$ for $J=J_{c}(h)$
 (Fig.\ref{fig:Bifurcation}(b)).
 As $J$ increases from $J_{c}(h)$, $z_{t}$ splits into $z_{-}$ and $z_{u}$.
 The change from $J<J_{c}(h)$ to $J>J_{c}(h)$ is discontinuous.
 This difference suggests that the phase transition is continuous for
 $h=0$ and discontinuous for $h>0$.

\section{\label{sec:cor}Asymptotic behavior of $C(t)$}
In this section, we derive the asymptotic form of 
the correlation function $C(t)$ using the previous results for 
$J\neq J_{c}(h)$. Based on them, we assume the functional form
 of $C(t)$ 
for $J=J_{c}(h)$. $C(t)$ is defined 
as the covariance between $X(1)$ and $X(t+1)$ 
divided by the variance of $X(1)$,$\mbox{Var}(X(1))$:
\[
C(t)\equiv \mbox{Cov}(X(1),X(t+1))/\mbox{Var}(X(1)).
\]
Through normalization, $C(0)=1$.
$C(t)$ can be expressed as the  difference 
between two conditional probabilities.
\begin{equation}
C(t)=\mbox{Pr}(X(t+1)=1|X(1)=1)-\mbox{Pr}(X(t+1)=1|X(1)=0) 
\label{eq:Ct}.
\end{equation}
 In general, $C(t)$ is positive for $J>0$.

\subsection{$C(t)$ for $J\neq J_{c}(h)$}
 The asymptotic behavior of $C(t)$ depends on $(J,h)$.
 If $J<J_{c}(h)$, there is one stable fixed point at $z_{+}$ 
 and $z(t)$ converges to $z_{+}$ through the  power-law
 relation $\mbox{E}(z(t)-z_{+})\propto t^{l_{+}-1}$ \cite{Hod:2004}.
 Here, the expectation value E$(A)$ of a certain quantity $A$ is defined as
 the ensemble average over the paths of the stochastic process. 
 If $J>J_{c}(h)$, there is another stable fixed point at $z_{-}$.
 Both $z_{\pm}$ are stable, and $z(t)$ converges to one of the 
 fixed points. The convergence of $z(t)$ to $z_{\pm}$ also 
 exhibits a power-law behavior 
 $\mbox{E}(z(t)-z_{\pm})\propto t^{l_{\pm}-1}$ \cite{His:2012}. 
 We assume that the probability that $z(t)$ converges to
 one of the $z_{\pm}$ depends on $X(1)$ and  we denote this as 
\[
p_{\pm}(x)\equiv \mbox{Pr}(z(t)\to z_{\pm}|X(1)=x).
\]
 For $J<J_{c}(h)$, $z(t)$ always converges to $z_{+}$ irrespective 
 of the value of $X(1)=x$, and  
$p_{+}(x)=1$ holds. In this case, we set
$p_{-}(x)=0$.
 Regarding the asymptotic behavior of the convergence 
 of $z(t)\to z_{\pm}$, which also depends on $X(1)$,
 we assume  
\[
\mbox{E}(z(t)\to z_{\pm}|X(1)=x)\simeq W_{\pm}(x)t^{l_{\pm}-1}
\]
 We write the dependence of the coefficients $W_{\pm}(x)$
 on the value of $X(1)$ explicitly.
 Using these behaviors and notations, we estimate the asymptotic 
 behavior of $C(t)$ as  
\begin{eqnarray}
C(t)&=& \mbox{Pr}(X(t+1)=1|X(1)=1)-\mbox{Pr}(X(t+1)=1|X(1)=0)
 \nonumber \\
&=& 
\mbox{E}(q(z(t))|X(1)=1)-\mbox{E}(q(z(t))|X(1)=0) \nonumber \\
&=&  
\sum_{x=0}^{1}(-1)^{x-1}
\left\{\mbox{E}(q(z(t))|x)
\mbox{Pr}(z(t)\to z_{+}|x)
+
\mbox{E}(q(z(t))|x)
\mbox{Pr}(z(t)\to z_{-}|x)  \right\}\nonumber \\
&\simeq &
\sum_{x=0}^{1}(-1)^{x-1} \left\{
(q_{+}+l_{+}\mbox{E}(z-z_{+}|x))
p_{+}(x)
+(q_{-}+l_{-}\mbox{E}(z-z_{-}|x))
p_{-}(x) \right\}\nonumber \\
&= &
\sum_{x=0}^{1}(-1)^{x-1}\left\{ 
(q_{+}+l_{+}W_{+}(x)t^{l_{+}-1})
p_{+}(x)
+
(q_{-}+l_{-}W_{-}(x)t^{l_{-}-1})
p_{-}(x) \right\} \nonumber \\
&=&
\sum_{x=0}^{1}
\left[q_{+}p_{+}(x)+q_{-}p_{-}(x)+
(l_{+}W_{+}(x)p_{+}(x)t^{l_{+}-1}+
l_{-}W_{-}(x)p_{-}(x)t^{l_{-}-1})\right](-1)^{x-1} .\nonumber  \\
\label{eq:c_t2}
\end{eqnarray}
Here, we expand $q(z)$ as
\[
q(z)=q(z_{\pm}+l_{\pm}\cdot (z-z_{\pm}))\simeq q_{\pm}+l_{\pm}\cdot (z-z_{\pm}).
\]
Given that $p_{+}(x)+p_{-}(x)=1$ for $x=0,1$, 
the limit value $c\equiv \lim_{t\to\infty}C(t)$ is estimated to be
\begin{equation}
c=(q_{+}-q_{-})(p_{+}(1)-p_{+}(0)). \label{eq:c}
\end{equation}
For $J<J_{c}(h)$, $p_{+}(x)=1$ and $c=0$.
As $z_{-}$ is stable for $J>J_{c}(h)$, the probability for the convergence 
of $z(t)$ to $z_{-}$ is positive.  
It is natural to assume that $p_{+}(1)>p_{+}(0)$ 
and $c>0$ for $J>J_{c}(h)$.

The asymptotic behavior of $C(t)$ is
 governed by the term with  
the largest value among $\{l_{+},l_{-}\}$ for $J>J_{c}(h)$.
We define $l_{max}$ as
\begin{equation} 
l_{max} \equiv 
\begin{cases}
l_{+} \,\,\,\, , \,\,\,\, \mbox{$J<J_{c}(h)$},   \\
\mbox{Max}\{l_{+},l_{-}\} \,\,\,\,  , \,\,\,\, \mbox{$J>J_{c}(h)$}.   
\end{cases}
\label{eq:l_max}
\end{equation}
We summarize the asymptotic behavior of $C(t)$ as
\begin{equation}
C(t)\simeq c+c'\cdot t^{l-1} \,\,\, \mbox{and}\,\,\, l=l_{max}. \label{eq:c_t} 
\end{equation}
Here, we write the coefficient of the term proportional to $t^{l-1}$ as $c'$.
If $J>J_{c}(h)$, the constant term $c$ is the leading term.
If $J<J_{c}(h)$, the power law term $c'\cdot t^{l-1}$ is the leading term. 
There also exists a sub-leading term to $c'\cdot t^{l-1}$ that we do not write
explicitly. One reason for this is that we do not understand 
the asymptotic behavior.
 The second reason is that our interest is focused on the value of $l$. 
 
\subsection{$J=J_{c}(h)$}

On the boundary $J=J_{c}(h)$, there are two stable points $z_{+}$ and 
$z_{t}$ for $h>0$. 
As $z_{t}$ is stable, the probability for the convergence 
of $z(t)$ to $z_{t}$ is positive.  
It is natural to assume that $p_{+}(1)>p_{+}(0)$ and $c>0$.
If $h=0$, there is only one stable point at $z_{+}=z_{t}=1/2$ and $c=0$.
As $l_{max}=l_{t}=1$, we anticipate that $|C(t)-c|$ 
becomes a decreasing function of $\ln t$. One possibility  
 is a power-law behavior of $\ln t$ such as
\begin{equation}
C(t)\simeq c+c'\cdot (\ln t)^{-\alpha'}.  \label{eq:c_logt} 
\end{equation}
In the case of the digital model,
 $C(t) \propto t^{-\alpha}$ with $\alpha=1/2$
for $p=p_{c}$.
We denote the power law exponent for $\ln t$
as $\alpha'$. 

 We derive $\alpha'$ by a simple heuristic argument. 
At first, we consider the case of $h=0$.
There is only one stable touchpoint at $z_{t}$, and 
$z(t)$ converges to it. 
Eq.(\ref{eq:c_t2}) suggests that 
the asymptotic behavior of $C(t)$
 is governed by $\mbox{E}(z-z_{t}|x)$ 
as $z_{t}$ is the only stable state.
As $q(z_{t})=z_{t}$ and $q'(z_{t})=1$ at $z_{t}$, 
$q(z)$ can be approximated in the vicinity of 
$z_{t}$ as
\[
q(z)=-\delta (z-z_{t})^{3}+z.
\]
Here $\delta$ is a positive constant, as 
$z_{t}$ is stable (Fig.\ref{fig:Bifurcation}a). 
The time evolution of $\mbox{E}(z-z_{t}|x)$ is given as
\[
\mbox{E}(z(t+1)-z_{t}|x)-\mbox{E}(z(t)-z_{t}|x)
=\frac{1}{t+1}\mbox{E}(q(z(t))-z(t)|x) 
\simeq -\frac{\delta}{t}\mbox{E}((z-z_{t})^{3}|x).
\]
Here the denominator $t+1$ in the middle of the equation 
comes from the fact that there occurs a $\frac{1}{t+1}$ change 
in E$(z(t)|x)$ for 
$X(t+1)\in \{0,1\}$.
We also assume $\mbox{E}((z(t)-z_{t})^{3}|x)\simeq 
\mbox{E}(z(t)-z_{t}|x)^{3}$ and 
the equation can be written as
\[
\frac{d}{dt}\mbox{E}(z(t)-z_{t}|x)=-\frac{\delta}{t}\mbox{E}(z(t)-z_{t}|x)^{3}.
\]
The solution to this shows the next asymptotic behavior
\[
\mbox{E}(z-z_{t}|x) \propto (\ln t)^{-1/2},
\]
 and we obtain $\alpha'=1/2$.

Likewise, for $h>0$, there are two stable states $q_{+}$ and $q_{t}$.
The subleading term 
in $C(t)$ is governed by $\mbox{E}(z-z_{t}|x)$. 
We can approximate $q(z)$ in the vicinity of $z_{t}$
to be
\[
q(z)=\delta(z-z_{t})^{2}+z.
\]
Here $\delta$ is a positive constant (Fig.\ref{fig:Bifurcation}b). 
If $z(t)>z_{t}$, $z(t)$ moves toward the right-hand direction, 
on average, and converges to $z_{+}$. 
We only  need to consider the case $z(t)<z_{t}$ and $z(t)$ converges 
to $z_{t}$. In the case,
E$(z(t)-z_{t}|x)$ obeys the next differential equation.
\[
\frac{d}{dt}\mbox{E}(z(t)-z_{t}|x)=\frac{\delta}{t}\mbox{E}(z(t)-z_{t}|x)^{2}.
\]
The solution shows the next asymptotic behavior
\[
\mbox{E}(z(t)-z_{t}|x)\propto (\ln t)^{-1},
\]
 and we obtain $\alpha'=1$.

\subsection{Numerical check of $C(t)\simeq c+c'\cdot t^{l-1}$}
We perform the numerical integration of the master equation of the system  
and check the asymptotic forms for $C(t)$ numerically.  
We denote the joint probability function for 
$\sum_{s=1}^{t}X(s)$ and $X(1)$ as
$P(t,n,x_{1})\equiv \mbox{Pr}(\sum_{s=1}^{t}X(s)=n,X(1)=x_{1})$.
For $t=1$, $P(1,1,1)=q(1/2)$ and $P(1,0,0)=1-q(1/2)$. 
The other components are equal to zero.
The master equation for $P(t,n,x_{1})$ is
\begin{equation}
P(t+1,n,x_{1})=q((n-1)/t)\cdot P(t,n-1,x_{1})
+(1-q(n/t))\cdot P(t,n,x_{1}).
\end{equation}
We impose the boundary conditions 
$P(t,n,x_{1})=0$ for $n<0$ or $n>t$.
Using $P(t,n,x_{1})$ for the case when $t\le T$, 
we estimate $C(t)$ for $t< T$.

\begin{figure}[htbp]
\begin{center}
\includegraphics[width=10cm]{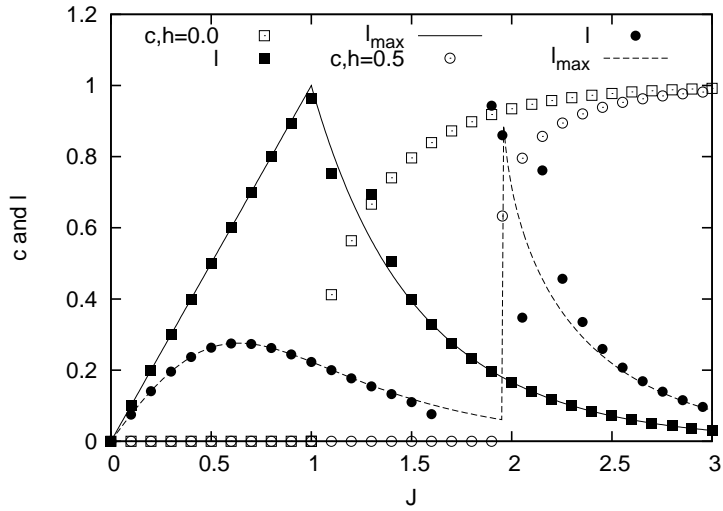}
\end{center}
\caption{
Plot of $c$ and $l$ in Eqs.(\ref{eq:cal}) and (\ref{eq:al}) for 
$h\in \{0.0,0.5\}$. 
$\Box (\blacksquare)$: $c (l)$ for $h=0$, 
 and $\circ (\bullet)$: $c (l)$ for $h=0.5$. 
 The solid (broken) line shows $l_{max}$ for $h=0.0 (0.5)$.
}
\label{fig:c_ll}
\end{figure}	

We adopt $h\in \{0.0,0.5\}$ and $J\in [0,3]$, and 
estimate $C(t)$ for $t< T= 4\cdot 10^{5}$.
The threshold values $J_{c}(h)$ are $J_{c}(0)=1$ and $J_{c}(0.5)\simeq 1.953$.
We assume the asymptotic behavior $C(t)\simeq c+c'\cdot t^{l-1}$ for 
$J >J_{c}(0),h=0$ and $J\ge J_{c}(h),h>0$. 
Using the three values of $C(t)$ at 
$t_{0}=T, t_{1}=T/s$, and $t_{2}=T/s^{2}$ with $s=2$, we solve for 
$c,c'$ and $l$ in the following manner.
\begin{eqnarray}
l&=&1-\ln_{s}\frac{C(t_{2})-C(t_{1})}{C(t_{1})-C(t_{0})}, \nonumber \\
c'&=&\frac{C(t_{1})-C(t_{0})}{t_{1}^{l-1}-t_{0}^{l-1}}, \nonumber \\
c&=&C(t_{0})-a\cdot t_{0}^{l-1} \label{eq:cal}.
\end{eqnarray}
For $J\le J_{c}(0),h=0$ and $J< J_{c}(h),h>0$, 
we assume 
the asymptotic behavior $C(t)\simeq c'\cdot t^{l-1}$.
In this case, we estimate $l,c'$ to be 
\begin{eqnarray}
l&=& 1-\ln_{s}C(t_{1})/C(t_{0}), \nonumber \\
c'&=&C(t_{0})\cdot t_{0}^{1-l} \label{eq:al}.
\end{eqnarray} 
$c$ and $l$ are plotted using symbols 
in Fig. \ref{fig:c_ll}. We also plotted $l_{max}$ of Eq.(\ref{eq:l_max})
with solid and broken 
lines. The estimation of $l$ obtained using Eqs.(\ref{eq:cal}) 
and (\ref{eq:al}) is nearly 
consistent with $l_{max}$.
However, for $1.2\le J\le 1.5,h=0.0$ and $2.1\le J\le 2.4,h=0.5$, 
large discrepancies are observed in the estimation of $l$. 
A possible explanation for these discrepancies is that the system 
size $T$ is not sufficiently large, and therefore the assumption of the aforementioned asymptotic behavior does not hold.
 In particular, as $t$ increases, the sign
of $c'$ changes at some particular $t$ value, 
 and this negatively affects the estimation of $l$. 
 We believe that these discrepancies 
 can be removed by increasing the system size $T$. 
 For $J=J_{c}(h)$, the 
 aforementioned fitting procedure does not 
 provide accurate results. 
 In the next section, we apply the FSS method 
 and clarify the asymptotic behavior 
 of $C(t)$ for $J=J_{c}(h)$.

\section{\label{sec:fss}Finite-size scaling analysis}
 The asymptotic behavior of a system 
 is governed by a temporal length scale, which is referred to as 
 the correlation length $\xi$. 
 For the definition of $\xi$, we adopt the 
 second moment correlation length of 
 $C(t)$\cite{Car:2000,Mor:2015}, which 
 has been adopted in the study of the equilibrium
 phase transition of spin models \cite{Bin:1985}
 and in the percolation theory \cite{Sta:1991}.
 We denote the $n-$th moment of $C(s)$ 
 for the period $s<t$ as 
 $M_{n}(t)\equiv \sum_{s=0}^{t-1}C(s)s^{n}$.
 The variable $t$ in $M_{n}(t)$ is considered as the time 
 horizon or system size of the stochastic process.  
 The second moment correlation length $\xi(t)$ is 
 defined as $\xi(t)\equiv \sqrt{M_{2}(t)/M_{0}(t)}$.
 The integrated correlation time, also referred to as the relaxation time, 
 $\tau(t)$, is defined as $\tau(t)\equiv M_{0}(t)$.
 $\xi(t)$ and $\tau(t)$ have the same dimensions; the length scale 
 in the critical behavior of the system is given by $\xi$. 

\subsection{FSS for $J \neq J_{c}(h)$}
We estimate $M_{n}(t)$ 
using the asymptotic behavior of $C(t)$ in Eq.(\ref{eq:c_t}).
\[
M_{n}(t)=\sum_{s=0}^{t}C(s)s^{n}\simeq \int_{0}^{t}C(s)s^{n}ds
= c\frac{t^{n+1}}{n+1}+c'\frac{t^{n+l}}{n+l}.
\]
$\xi_{t}(t)\equiv \xi(t)/t=\sqrt{M_{2}(t)/M_{0}(t)t^{2}}$ 
is then given as 
\begin{equation}
\xi_{t}(t)
\simeq
\begin{cases}
\sqrt{\frac{l}{l+2}}\,\,\,\, , \,\,\,\, c=0 \\
\sqrt{\frac{1}{3}}\left(1+\frac{c'}{2c}(\frac{3}{l+2}-\frac{1}{l})t^{l-1}\right)
\,\,\,\, , \,\,\,\,c>0. \label{eq:xit1}
\end{cases}
\end{equation}
$\xi_{t}$ converges to 
$\sqrt{1/3}$ and $\sqrt{l/(l+2)}$ for $c>0$ and $c=0$,
respectively. 

In the case $c>0$, we denote the deviation of $\xi_{t}$ from 
the limit vale $\sqrt{1/3}$ normalized by the limit value as $\Delta \xi_{t}$.
\[
\Delta \xi_{t}(t)\equiv \frac{\xi_{t}(t)-\sqrt{1/3}}{\sqrt{1/3}}
\]
$\xi_{t}(t)$ behaves as 
\[
\xi_{t}(t)
=\sqrt{\frac{1}{3}}(1+\Delta \xi_{t}(t)).
\]
We also describe $\tau_{t}(t)\equiv \tau(t)/t$ as
\begin{equation}
\tau_{t}(t)\simeq c+\frac{c'}{l}t^{l-1} \label{eq:taut1}. 
\end{equation}
We express $t^{l-1}$ using $\Delta \xi_{t}(t)$ for $c>0$ as 
\begin{equation}
\tau_{t}(t)=c\left(1+\frac{l+2}{l-1}\Delta \xi_{t}(t)\right). 
\end{equation}
$\tau_{t}$ converges to $c$ 
as $\Delta \xi_{t}(t)$ converges to zero. 

We define the scale transformation of the system 
as the change in the time 
horizon of the system $t\to s t$ with the scale factor $s$. 
We denote the scaling function for a long-term observable $A(t)$
 as $f_{A}$, which is defined as
\[
\frac{A(st)}{A(t)}=f_{A}(\xi_{t}(t))+\Delta f_{A}(t).
\] 
Here, we include the correction to the scaling term  as $\Delta f_{A}(t)$. 
$\Delta f_{A}(t)$ is of the order 
$t^{-\omega}$ and $\omega$ is a correction-to-scaling exponent\cite{Car:2000}.

When $C(t)$ exhibits power-law behavior, 
  the system is scale invariant \cite{Mor:2015}. 
Furthermore, $\lim_{t\to \infty}\xi_{t}(t)$ is constant and  $f_{\xi_{t}}=1$. 
$\Delta f_{\xi_{t}}$ for $c>0$ is given by  
\[
\Delta f_{\xi_{t}}=\frac{\xi_{t}(st)}{\xi_{t}(t)}-1\simeq 
(s^{l-1}-1)\Delta \xi_{t}.
\]
As $\Delta \xi_{t}\propto t^{l-1}$, $\omega=l-1$ for $\xi_{t}$.

If $c>0$, $f_{\tau_{t}}=1$ as $\lim_{t\to \infty}\tau_{t}=c$, then
\[
\Delta f_{\tau_{t}}=\frac{\tau_{t}(st)}{\tau_{t}(t)}-1\simeq 
\frac{l+2}{l-1}(s^{l-1}-1)\Delta \xi_{t}.
\]
and $\omega=l-1$.
If $c=0$, $\tau_{t}\propto t^{l-1}$ and 
\begin{equation}
\ln_{s}f_{\tau_{t}}=
\lim_{t\to \infty}\ln_{s}\frac{\tau_{t}(st)}{\tau_{t}(t)}=l-1=
\frac{3-\xi_{t}^{-2}}{\xi_{t}^{-2}-1}.
\end{equation}
As $l<1$, under the scale transformation 
$t\to st$, $\tau_{t}(st)=s^{l-1}\tau_{t}(t)\propto s^{l-1}$. 
In the limit $s\to \infty$, $\tau_{t}(st)$ decreases to zero. 

\subsection{FSS for $J=J_{c}(h)$}
We consider and verify Eq.(\ref{eq:c_logt}) 
by studying the finite-size scaling relation of the system. 
We estimate $M_{n}(t)$ to $t^{n+1}(\ln t)^{-\alpha'-1}$ as
\begin{eqnarray}
M_{n}(t)&=&\int^{t}C(s)s^{n}ds\simeq c\frac{t^{n+1}}{n+1}+c'\int^{t} s^{n}
(\ln s)^{-\alpha'}ds \nonumber \\
&\simeq &c\frac{t^{n+1}}{n+1}
+c'\frac{t^{n+1}}{n+1}(\ln t)^{-\alpha'}
\left(1+\frac{\alpha'}{n+1} (\ln t)^{-1}\right). \nonumber
\end{eqnarray}
Here, we use the next expansion in powers of $1/\ln t$, 
which is obtained by partial
 integration. 
\begin{eqnarray}
&&\int^{t} s^{n}
(\ln s)^{-\alpha'}ds=\frac{1}{n+1}t^{n+1}(\ln t)^{-\alpha'}
+
\frac{\alpha'}{n+1}\int^{t} s^{n}
(\ln s)^{-\alpha'-1}ds \nonumber \\
&=&\frac{1}{n+1}t^{n+1}(\ln t)^{-\alpha'}
+\frac{\alpha'}{(n+1)^{2}}t^{n+1}(\ln t)^{-\alpha'-1}
+\frac{\alpha'(\alpha'+1)}{(n+1)^{2}}\int^{t} s^{n}
(\ln s)^{-\alpha'-2}ds \nonumber \\
&=& \frac{1}{n+1}t^{n+1}(\ln t)^{-\alpha'}
+\frac{\alpha'}{(n+1)^{2}}t^{n+1}(\ln t)^{-\alpha'-1}
+O(t^{n+1}(\ln t)^{-\alpha'-2}). \nonumber 
\end{eqnarray}
In the following, we omit the parentheses $(,)$ in $(\ln t)^{-x}$
when we write the power of $\ln t$.
$\xi_{t}(t)$ is
\[
\xi_{t}\simeq
\begin{cases}
\sqrt{\frac{1}{3}}\left(1-\frac{2\alpha'}{3}\ln t^{-1}\right)
\,\,\,\, , \,\,\,\, c=0 \\
\sqrt{\frac{1}{3}}\left(1-\frac{c'}{3c}(\ln t^{-\alpha'-1}
-\frac{1}{4}\ln t^{-2\alpha'})\right)
 \,\,\,\, , \,\,\,\,c>0.
\end{cases}
\]
$\xi_{t}$ converges to the same value $\sqrt{1/3}$ in 
both cases, $c=0$ and $c>0$. However, 
the convergence speeds differ. 
The coefficients and exponents of the power of $\ln t$
of the sub-leading terms are given as $B$ and $\beta'$, respectively.
\begin{equation}
\xi_{t} \simeq 
\sqrt{\frac{1}{3}}(1-B \ln t^{-\beta'})
\equiv \sqrt{\frac{1}{3}}(1+\Delta \xi_{t}(t)) \label{eq:xit2}
\end{equation}
Using the second equality, we define $\Delta \xi_{t}$.
$\beta'=1$ for $c=0$ and  
$\beta'=\mbox{Min}(1+\alpha',2\alpha')$ for $c>0$.

$\tau_{t}(t)$ is estimated to be 
\begin{equation}
\tau_{t}\simeq c+c' \ln t^{-\alpha'} \label{eq:taut2}. 
\end{equation}

As $\xi_{t}$ converges to $\sqrt{1/3}$, the system is scale-invariant
 and $f_{\xi_{t}}=1$.  
For $c=0$, we can use the scaling relation for $\tau_{t}$ in order to 
estimate $\alpha'$.
Because $\tau_{t}$ is a  function of $\ln t$, we consider the exponential 
scaling transformation $t\to t^{s}$ with the scale factor $s$. 
\begin{equation}
\frac{\tau_{t}(t^{s})}{\tau_{t}(t)}=s^{-\alpha'} \label{eq:est_alpha}.
\end{equation}
For $c>0$, as $f_{\tau_{t}}=1$,
it is necessary to study the finite-size scaling correction for $\tau_{t}$.
\begin{equation}
\frac{\tau_{t}(t^{s})}{\tau_{t}(t)}-1\simeq \frac{c'}{c}(s^{-\alpha'}-1)
\ln t^{-\alpha'} \label{eq:est_alpha2}.
\end{equation}
We can estimate $\alpha'$ and $c'/c$  using Eq.(\ref{eq:est_alpha2}).
In addition, we can estimate $\beta'$ using 
$\xi_{t}$.
\begin{equation}
\frac{\xi_{t}(t^{s})}{\xi_{t}(t)}-1\simeq (s^{-\beta'}-1)\Delta \xi_{t}.
\label{eq:est_beta}
\end{equation}

If $\alpha'=1$, a convenient extrapolation formula for $c$ is available. 
As $\Delta \xi_{t}=-\frac{c'}{4c}(\ln t)^{-2}$, we can express $c'/c$
 using $\Delta \xi_{t}$ and $\ln t$. 
We obtain the extrapolation formula for $c$ by
solving Eq.(\ref{eq:taut2}) as
\begin{equation}
c=\frac{\tau_{t}(t)}{1-4\Delta \xi_{t}(t)\ln t} \label{eq:est_c_Jc}.
\end{equation}

\section{\label{sec:fss_Jc}Numerical studies of FSS and $C(t)$ for $J=J_{c}(h)$}

\begin{figure}[htbp]
\begin{center}
\begin{tabular}{c}
\includegraphics[width=9cm]{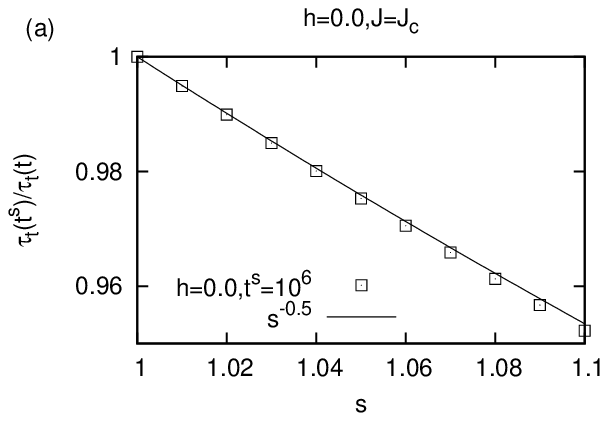} 
\vspace*{0.3cm} \\
\includegraphics[width=9cm]{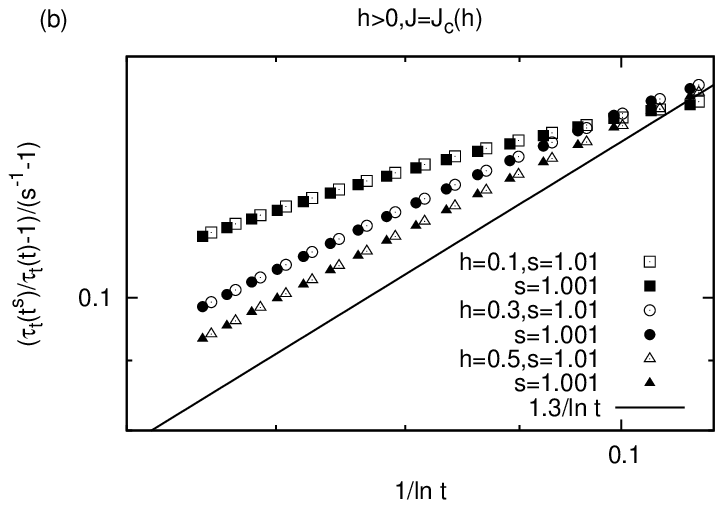} 
\vspace*{0.3cm} \\
\includegraphics[width=9cm]{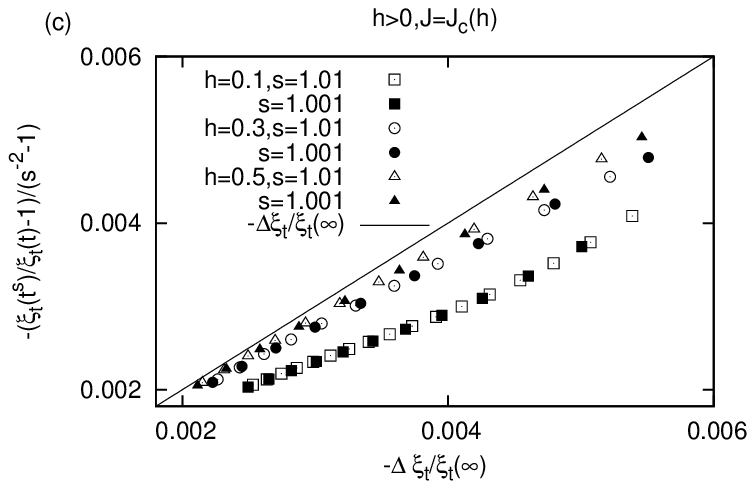} 
\end{tabular}
\end{center}
\caption{\label{fig:FSS_Jc}
(a) Plots of 
$\tau_{t}(t^{s})/\tau_{t}(t)$ vs. $s$.
(b) Plots of 
$(\tau_{t}(t^{s})/\tau_{t}(t)-1)/(s^{-1}-1)$ vs. $1/\ln t$.
(c) Plots of 
$-(\xi_{t}(t^{s})/\xi_{t}(t)-1)/(s^{-2}-1)$ vs. $-\Delta \xi_{t}(t)$. 
$t<T=5\times 10^{6} (3\times 10^{6} )$ for $h>0 (h=0)$
}
\end{figure}

We check the assumption $C(t)\simeq c+c'\ln t^{-\alpha'}$ 
by studying finite-size scaling.
First, we estimate $\alpha'$ for $h=0$ using  
Eq.(\ref{eq:est_alpha}) and the data
for $t^{s}=3\times 10^{6}$ and $s\in [1.0,1,1]$.
Figure \ref{fig:FSS_Jc}(a) 
shows the plotted results.
It is evident that the data lies on the curve $s^{-1/2}$ and
 $\alpha'=1/2$.
 
 For $h>0$, we apply Eq.(\ref{eq:est_alpha2}) in order 
to estimate $\alpha'$. We estimate 
$\tau_{t}(t^{s})/\tau_{t}(t)-1$ for $s=1.01,1.001$
and divide it by $s^{-\alpha'}-1$ using the data for $h\in \{0.1,0.3,0.5\}$
 and $t^{s}\le 5\times 10^{6}$.
We choose $\alpha'$ so that two data sets with different $s$
lie on the same curve as a function of $1/\ln t$. 
Figure \ref{fig:FSS_Jc}(b)
shows the double lnarithmic plot of the obtained results. 
We choose $\alpha'=1$ and 
two data sets for each $h$ that lies 
on the same curve. If the curve obeys Eq.(\ref{eq:est_alpha2}),
 the slope should be equal to 1 and then, we can estimate $c'/c$ by 
 fitting with $c'/c\ln t$.
 However, even for $t^{s}=5\times 10^{6}$, the slope of the curve 
 is not equal to one for $h=0.1$.
 We suppose that the system size $t^{s}=5\times 10^{6}$ is not large enough for  
 Eq.(\ref{eq:est_alpha2}) to hold in this case.  

In order to verify that $\alpha'=1$ for $h>0$, 
we use Eq.(\ref{eq:est_beta}).
As $\beta'=\mbox{Min}(1+\alpha',2\alpha')$, $\beta'=2$ for $\alpha'=1$.
We use the estimate of $\xi_{t}(t^{s})/\xi_{t}(t)-1$ with the same data set and 
$s=1.01,1.001$. We divide it by $s^{-2}-1$. Figure 
\ref{fig:FSS_Jc}(c) shows the
plotted results. The two data sets for each $h$
 lie on the same curve; this agrees with our estimation $\beta'=2$. 
Furthermore, as $t$ becomes large, the data set converges to $\Delta \xi$.  
 These results are consistent with Eq.(\ref{eq:est_beta}) and 
confirm that $\alpha'=1$. 

\begin{figure}[htbp]
\begin{center}
\includegraphics[width=10cm]{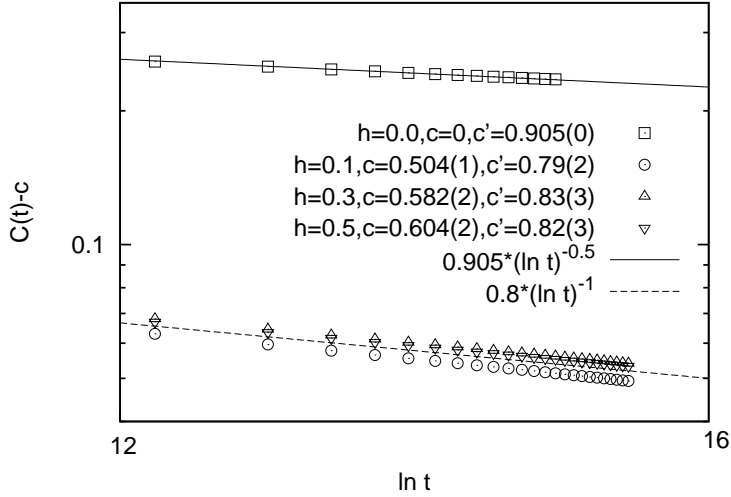} 
\end{center}
\caption{\label{fig:Cor_Jc}
Plots of $C(t)-c$ vs. $\ln t$ for $J=J_{c}(h)$ and $h\in \{0.0,0.1,0.3,0.5\}$. 
$t<T=5\times 10^{6} (3\times 10^{6} )$ for $h>0 (h=0)$.
 For $h=0.0$, we plot the fitted results 
using  $C(t)=c'\ln t^{-1/2}$ with a fitting parameter of $c'$. 
For $h>0.0$, we plot the fitted results using $C(t)=c+c'\ln t^{-1}$ 
with two fitting parameters $c,c'$.
}
\end{figure}

We estimate the fitting parameters $c,c'$ using the numerical
results of $C(t)$ for $t\le 5\times 10^{6}$.
After $c$ is subtracted from $C(t)$, $C(t)-c$ behaves as 
$c'(\ln t)^{-1}$ for $c>0$. Figure \ref{fig:Cor_Jc} shows a 
double lnarithmic plot of $C(t)-c$ vs. $\ln t$ for $J=J_{c}(h)$.
 It is difficult to verify the power law dependence of $C(t)-c$ 
 on $\ln t$  using the narrow range of $\ln t$. The data 
 lies on the straight lines.  

\subsection{Extrapolation of $c$}
We now estimate the limit value $c=\lim_{t\to\infty}C(t)$ for $J=J_{c}(h)$.
A simple method is to use the fitted result with the assumption
that $C(t)=c+c'\ln t^{-1}$. Another method is to use Eq.(\ref{eq:est_c_Jc}).
We summarize the results of these methods in Table \ref{tab:c_Jc}.
\begin{table}[htbp]
\begin{center}
\caption{\label{tab:c_Jc}
The columns in the table denote the following: an estimation of $c$ for $h\in \{0.1,0.3,0.5\}$ and $J=J_{c}(h)$. $h$ in the first column;
$C(T-1)$ in the second column; the estimate by fitting
in the third column; $\tau_{t}(T)$ in the forth column;
the estimate using Eq.(\ref{eq:est_c_Jc}) 
in the fifth column. $T=5\times 10^{6}$.}
\begin{tabular}{ccccc}
\hline
h & $C(T-1)$ & $c$ (Fitted) & $\tau_{t}(T)$  & $c$ (Extrapolated)   \\ 
\hline
0.1 & 0.558  & 0.504(1) & 0.560  & 0.514    \\ 
0.3 & 0.638   & 0.582(2) & 0.640  & 0.593  \\ 
0.5 & 0.659  & 0.604(2) & 0.661  & 0.615  \\ 
\hline
\end{tabular}
\end{center}
\end{table}

Both $C(T-1)$ and $\tau_{t}(T)$ provide slightly larger estimates of $c$
 than the fitting and extrapolation methods.
The fitted and extrapolated values are approximately 10\% smaller than the estimates using  $C(T-1)$ and $\tau_{t}(T)$ for $T=5\times 10^{6}$. An extremely slow 
convergence of $C(t)$ to $c$ is observed.

\section{\label{sec:UC}Universality class}

The system shows a continuous phase transition for $h=0$.
As we have seen, the analysis does not depend on the 
precise form of $q(z)$. As far as $q(z)$ has $Z_{2}$-symmetry, 
$q(1-z)=1-q(z)$ and $y=q(z)$ are tangential to $y=z$ at $z_{t}=1/2$, 
 we can assume a similar continuous phase transition. 
In order to  discuss the universality class of the phase transition,
 we compare the scaling properties of $C(t)$ for two models.
One model is the logistic model which adopts the $q(z)$ 
in Eq.(\ref{eq:model}). For $h=0$, $J_{c}(0)=1$.
Another model adopts the next $q_{r}(z)$ with 
three parameters $r,p\in [0,1],q_{*}\in (1/2,1]$\cite{His:2015a}.
\begin{eqnarray}
q_{r}(z)&=&(1-p)\cdot q_{*}+p\cdot \pi_{r}(z)  \nonumber \\
\pi_{r}(z)&=&\sum_{s=(r+1)/2}^{r}{}_{r}C_{s}\cdot z^{s}(1-z)^{r-s} \label{eq:model2} 
\end{eqnarray}
Here ${}_{r}C_{s}$ is the binomial coefficient and $r$ is a 
odd number greater than three, $r\in \{3,5,7\cdots\}$.
In the limit $r\to \infty$, $q_{r}(z)$ reduces to that of the digital model,
 $(1-p)\cdot q_{*}+p\cdot \theta(z-1/2)$. This model corresponds to the 
 mean-field approximation of the model,
 where $X(t)$ chooses the majority of $r$ randomly chosen 
  previous variables with a probability of $p$. For $q_{*}=1/2$, $q_{r}(z)$ 
 has $Z_{2}$-symmetry and the threshold value $p_{c}(r)$ is 
 determined by the condition $1=q_{r}'(z_{t}=1/2)=p_{c}(r)\cdot \pi_{r}'(1/2)$. 
 $p_{c}$ is explicitly given as
\[
p_{c}(r)=\frac{[(r-1)/2!]^{2}2^{r-1}}{r!}.
\]
$p_{c}(r)$ is a decreasing function of $r$ and 
$\lim_{r\to \infty}p_{c}(r)=0$, which is 
compatible with $p_{c}=1-1/2q_{*}=0$ of the digital model with $q_{*}=1/2$. 

At $J=J_{c}(0)$, $C(t)$ obeys a power-law of $\ln t$
as $C(t)\simeq  c'\ln t^{-\alpha'}$ with $\alpha'=1/2$ for the former model.
Below $J_{c}(0)=1$, $C(t)\propto t^{l-1}$ with $l=l_{+}=J$.
We set $\Delta J=J_{c}-J=1-J$. The expression 
for the exponent of $C(t)\propto t^{1-l}$ is given by 
\[
1-l=\Delta J \,\,\,\, , \,\,\,\, J<J_{c}(0)=1.
\]
For $J>1$, $C(t)-c \propto t^{l-1}$ with $l=l_{+}=l_{-}=q'(z_{+})$.
As in the case of the estimation of the critical exponents for the 
mean-field Ising model \cite{Sta:1971}, we solve $z_{+}=q(z_{+})$ 
with the assumption $\Delta J\equiv J-1<<1$. 
We obtain $z_{+}-1=\sqrt{3\Delta J}$ and  estimate $l$
as
\[
1-l=\frac{1}{2}\Delta J\,\,\,\, , \,\,\,\, J>J_{c}(0)=1.
\]
As $J$ approaches $J_{c}(0)$ from below and above of $J_{c}$, 
$1-l$ approaches 0. At $J=J_{c}$, $C(t)$ obeys a power-law of $\ln t$.
As $t^{-(1-l)}=e^{-(1-l)\ln t}$, we can regard $1/(1-l)$ as the 
"correlation length" while assuming $C(t)$
 as function of $\ln t$. We assume the phenomenolnical 
 scaling ansatz for $C(t)$ to be  
\begin{equation}
C(t)=\ln t ^{-\alpha'}g((1-l)\ln t) \label{eq:universal}.
\end{equation} 
In the ansatz, $\ln t$ dependence of $C(t)$
is scaled with $1/(1-l)$. $g(x)$ is the universal function and 
is finite at $x=0$. For $J=J_{c}$ and $l=1$, 
$C(t)=g(0)\ln t ^{-\alpha'}$ with $g(0)=c'$. In the limit $x\to \infty$,
in order to compensate the $\ln t^{-\alpha'}$ term, $g(x)$ should behave 
as $g(x)\propto x^{\alpha'}$ for $J>J_{c}(0)$. Then $C(t)$ behaves as
$\lim_{t \to \infty}C(t)\propto (1-l)^{\alpha'}=\Delta J^{\alpha'}$.
The critical exponent $\beta$ for $c\propto \Delta J^{\beta}$ 
coincides with $\alpha'$.
The exponent $\nu_{||}$ is defined for the divergence 
 of $1/(1-l)$ as 
\[
1/(1-l)\propto \Delta J^{-\nu_{||}}.
\]
As $(1-l)\propto \Delta J$, 
we obtain $\nu_{||}=1$.
 The scaling relation $\beta=\alpha'\cdot \nu_{||}$
 holds.

For the latter model with $q_{r}(z)$, $l=p\cdot q'_{r}(z_{t}=1/2)$ for 
$p<p_{c}(r)$. As $1=p_{c}\cdot q_{r}'(1/2)$ holds,  
the correlation length $1/(1-l)$ is estimated to be
\[
1/(1-l)=\frac{1}{\pi'_{r}(1/2)(p_{c}(r)-p)}
\]
and diverges as $1/(1-l)\propto \Delta p^{-\nu_{||}}$ with $\nu_{||}=1$. 
Here, we define $\Delta p\equiv p_{c}(r)-p$.
For $p>p_{c}$, we assume $\Delta p=p-p_{c}(r)<<1$ and estimate $z_{+}$ by solving 
$z_{+}=q_{r}(z_{+})$ up to $O(\Delta z^{3})=O((z_{+}-1/2)^{3})$.
One may then show that $1-l=q_{r}'(z_{+})\propto \Delta p$, 
and we obtain $\nu_{||}=1$. 
If we assume the scaling form for $C(t)$ to be
$C(t)\simeq \ln t^{-\alpha'}\cdot g(\ln t \cdot (1-l))$ and define $\beta$ as 
$c\propto \Delta p^{\beta}$, we obtain $\beta=\alpha'\cdot \nu_{||}$.

If the two models share the same value for $\alpha'$, 
this suggests that they are in the same universality class.

\subsection{Numerical calculation of $g(x)$}
 We estimate the universal function $g(x)$ assumed in Eq.(\ref{eq:universal})
 numerically. 
For the former logistic model with $h=0$, we estimate that $\alpha'=1/2$.
$J_{c}(0)=1$ and we estimate $C(t)$ for $t\le 4\times 10^{5}$ and 
$2/3<J<3/2$.
For the latter model with $r=3$ and $q_{*}=1/2$, $p_{c}(3)=2/3$.
We estimate $C(t)$ for $t\le 4\times 10^{5}$ and $4/9<p<1$.
Using data for $C(t)$ between  $10^{4}\le 
t\le 4\times 10^{5}$, we determine $g(x)$ to be
\[
g(x)=\ln t^{1/2}\cdot C(t)\,\,\,\, , \,\,\,\,  x=(1-l)\ln t.
\]
$g(x)$ should be smooth near $x=0$ and $g(0)=c'$
For a sufficiently large $J$, $c\simeq 1$ and $l=0$.
$g(x)$ should behave as $x^{1/2}$ for sufficiently large $x$ values.
For $J<1$, $g(x)$ should decrease exponentially.

\begin{figure}[htbp]
\begin{center}
\begin{tabular}{c}
\includegraphics[width=12cm]{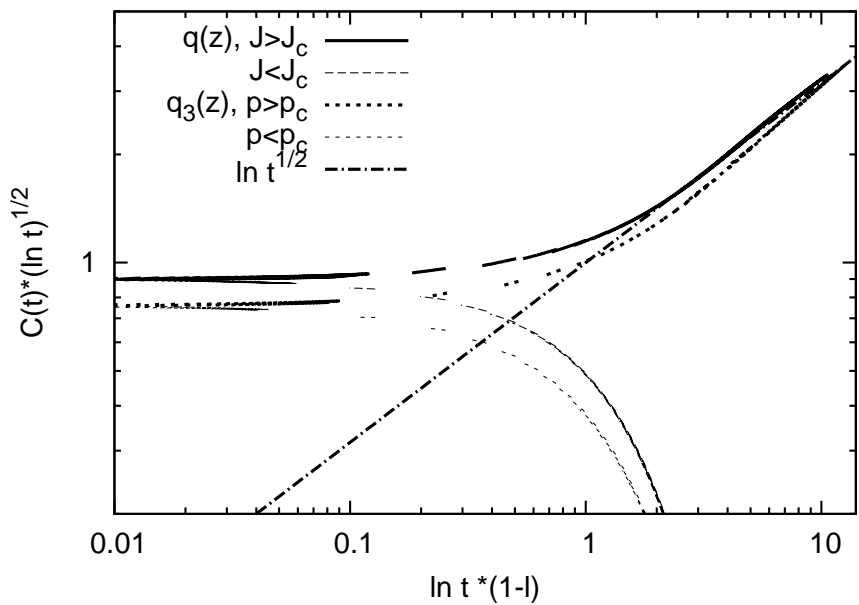} 
\end{tabular}
\end{center}
\caption{\label{fig:UC_I_r3}
Plot of $\ln t^{1/2}\cdot C(t)$ vs. $(l-1)\ln t$.
}
\end{figure}

Figure \ref{fig:UC_I_r3} shows the results of this analysis. 
The thick continuous and thick dashed lines indicate the results
for $J>J_{c}(0)$ and $p>p_{c}(3)$, respectively.
As can be clearly seen in this figure, the data obtained for
 different $J$ and $t$ values
and for different $p$ and $t$ values
lie on two curves, 
which represent $g(x)$ in the phase with $c>0$ for both models.
For large $x$, $g(x)\simeq x^{1/2}$.
 The thin continuous and thin dashed lines 
 indicate the results for $J<J_{c}(0)$ and $p<p_{c}(3)$, respectively.
 The data lie on two curves, which represent 
 $g(x)$ in the phase with $c=0$ for both models.
 $g(x)$ can be seen to decay exponentially. 
 The results indicate that the scaling ansatz in 
 Eq.(\ref{eq:universal}) holds with $\alpha'=1/2$.

\section{\label{sec:con}Summary and Notes}

In this study, we analyzed the asymptotic behavior of 
the normalized correlation function $C(t)$ for a generalized
P\'{o}lya urn. The probability $q(z)$ of adding a 
red ball to a specific proportion of red balls $z$  
is $q(z)=(\tanh [J(2z-1)+h]+1)/2$. There are three domains
 in $(J,h)$: $J<J_{c}(h)$, $J>J_{c}(h)$, and $J=J_{c}(h)$,
 where for $J=J_{c}(h)$, $q(z)$ becomes tangential to $z$.
 The limit value $c=\lim_{t\to\infty}C(t)$ is the order parameter
 for the phase transition. If $J<J_{c}(h)\,\,\,(>J_{c}(h))$,$c=0\,\,\,(c>0)$.
 $C(t) \sim c+a\cdot t^{l-1}$ with $c=0\,(>0)$ for $J<J_{c}\,(J>J_{c})$, 
 and $l$ is the (larger) value of the slope(s) of $q(z)$ at  
 stable fixed point(s).  Through FSS analysis, we evaluated the 
 asymptotic behavior of $C(t)$ for $J=J_{c}(h)$ as
 $C(t)\simeq c+c'\ln(t)^{-\alpha'}$.
 For $h=0$, $c=0$ and $\alpha'= 0.5$.
 For $h\neq 0$, $c>0$ and $\alpha'= 1$.
 The system shows a continuous phase transition for $h=0$.
$C(t)$ behaves as $C(t)=\ln t^{-\alpha'}g((1-l)\ln t)$ with a universal 
function and we numerically estimated $g(x)$. The scaling relation 
$\beta=\alpha'\nu_{||}$ holds true among the critical exponents 
with $\alpha'=1/2,\nu_{||}=1$ and $\beta=1/2$.
We also studied the scaling of $C(t)$ for 
$q_{r}(z)$ given in Eq.(\ref{eq:model2}) with $z_{*}=1/2$ and $r=3$.
 We showed that the scaling ansatz $C(t)=\log t^{-\alpha'}g((1-l)\log t)$
 holds for $\alpha'=1/2$.

 We note several key points and future challenges.
 The first of these is related to the relationship between the non-equilibrium 
 phase transition 
 studied in this work and equilibrium phase transition of the mean-field 
 Ising (MFI) model for $h=0$.
 As is well known, $\beta=1/2$ for the latter model 
 and both types of phase transitions give the same values for 
 $\beta$. A crucial difference lies in the behavior of $C(t)$.
 In the equilibrium phase transition, the memory of the value of $X(1)$
 should disappear for $J<J_{c}(0)$. However, in the case of the 
 non-equilibrium transition, 
 $C(t)$ decays according to a power-law of $t$ and continues to exist for finite $t$.
 Furthermore, for $h>0$, phase transition does not occur in the 
 equilibrium case and $z(t)$ converges to $z_{+}$.
 In the non-equilibrium case, the probability for
 $z(t)$ to converge to $z_{-}$ is positive for $J>J_{c}(h)$.
 In our previous study, we controlled the length of the memory $r$, and also,
 $X(t+1)$ depended only on recent $r$ variables $X(t-r+1),\cdots, X(t)$ 
 \cite{His:2015}.
 The P\'{o}lya urn process corresponds to the case where $r=t$. 
 We have shown that with the logarithmic increase  
 of $r\propto \ln t$, the non-equilibrium phase transition for $h>0$ 
 disappears and $z(t)$ always converges to $z_{+}$. 
 We believe that it is possible to 
 understand the relation between the non-equilibrium and equilibrium 
 phase transitions by controlling the increase of $r$.
 We assert that if the increase in $r$ is infinitely slowly and 
 the variables are completely equilibrated among the recent $r$ variables,
 the non-equilibrium phase transition reduces to the 
 equilibrium phase transition.

 A problem for the future is related to 
 the derivation of $\alpha'$ and $g(x)$.
 Our study only provides numerical results
 and a heuristic derivation of $\alpha'$;
 A more rigorous mathematical treatment appears to be necessary.
 As the model with $q_{r}(z)$ shares the same value of $\alpha'=1/2$ for $r=3$,
 and the heuristic derivation of $\alpha'$ only uses the approximate form 
 of $q(z)$ in the vicinity of the touchpoint,
 the universality class of the continuous phase transition of 
 a generalized P\'{o}lya, where $q(z)$ has $Z_{2}$-symmetry and 
 $q(z)$ becomes tangential to the diagonal at 
 $z_{t}=1/2$, is described by $\alpha'=1/2$. 

 It is also important to verify these results 
 using experimental data. Information cascade experiments could
 be one possible solution\cite{Mor:2012,Mor:2013}. 
As the empirically 
 estimated $q(z)$ does not
 exhibit $Z_{2}$-symmetry at $p=p_{c}$, the results of this paper 
 suggests that the transition in the experiment is 
 discontinuous. 
 The system size $T$ is severely limited in laboratory experiments. 
 Therefore, web-based experimental 
 systems 
should be developed in order to study the asymptotic 
 behavior of the system \cite{Sal:2006}.

\begin{acknowledgments}
This work was supported by Grant-in-Aid for Challenging 
Exploratory Research 25610109. 
\end{acknowledgments}

\bibliography{myref}

\appendix

\section{$g(x)$ for the "digital" model}
We derive the universal function $g(x)$ for the exactly solvable digital
model using the results given in Ref.\cite{Mor:2015}.
We adopt $q(z)=(1-p)\cdot q_{*}+p\cdot \theta(z-1/2)$.
The model shows a continuous phase transition at $p=p_{c}(q_{*})=1-1/2q_{*}$ 
for $q_{*}>1/2$. $C(t)$ behaves as $C(t)\simeq b(q_{*})t^{-1/2}$ at 
$p=p_{c}(q_{*})$. The limit value 
$c(q_{*},p)\equiv \lim_{t\to \infty}C(t)$  is a continuous function 
of $q_{*},p$ and becomes positive for $p>p_{c}(q_{*})$.

\begin{figure}[htbp]
\begin{center}
\begin{tabular}{c}
\includegraphics[width=12cm]{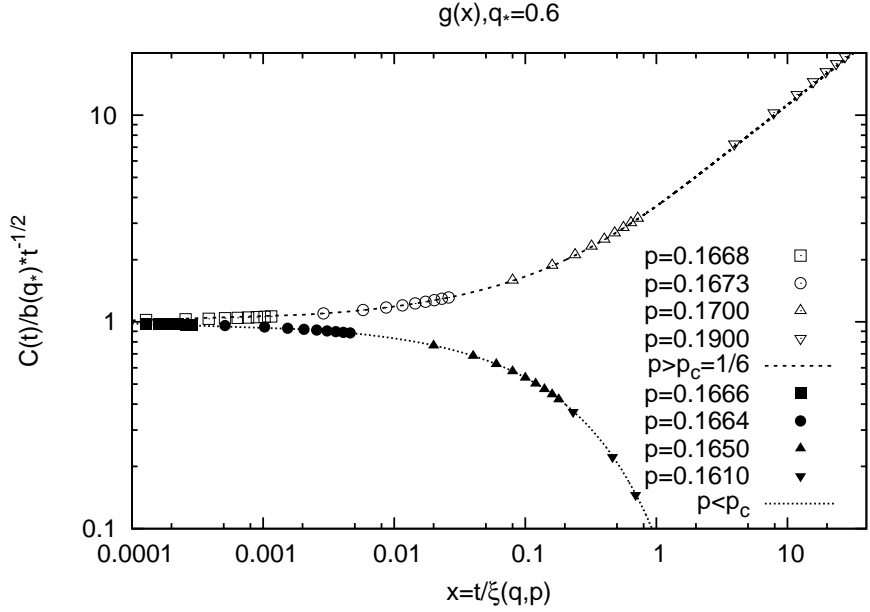} 
\end{tabular}
\end{center}
\caption{\label{fig:UC_D}
Plot of $C(t)/b(q_{*})t^{-1/2}$ vs. $t/\xi(q_{*},p)$ with empty symbols 
indicating $p>p_{c}(q_{*})=1/6$ 
and filled symbols indicating $p<p_{c}(q_{*})$.
We adopt $q_{*}=0.6$ and $10^{4}\le t\le 10^{5}$.
The lines show the results of Eq.(\ref{eq:UC_D}). 
}
\end{figure}

We assume the scaling form for $C(t)$ as
\begin{equation}
C(t)=b(q_{*})t^{-1/2}g(t/\xi(q_{*},p)) \label{eq:UC_D0}.
\end{equation}
$b(q_{*})$ and $\xi(q_{*},p)$ are defined as
\begin{eqnarray}
b(q_{*})&=&\sqrt{\frac{8}{\pi}}\left(\frac{2q_{*}-1}{4q_{*}-1}\right) \nonumber \\
\xi(q_{*},p)^{-1}&=&-\ln \sqrt{4(p+(1-p)(1-q_{*}))((1-p)q_{*})} \nonumber \\.
\end{eqnarray}
$g(x)$ is then given as
\begin{equation}
g(x)= 
\begin{cases}
\sqrt{4\pi x}+\frac{x^{1/2}}{2}\int_{x}^{\infty}u^{-3/2}e^{-u}du 
\,\,\,\, , \,\,\,\, \mbox{$p>p_{c}(q_{*})$},   \\
\frac{x^{1/2}}{2}\int_{x}^{\infty}u^{-3/2}e^{-u}du
 \,\,\,\,  , \,\,\,\, \mbox{$p<p_{c}(q_{*})$}.   
\end{cases}
\label{eq:UC_D}
\end{equation}
For the derivation of $g(x)$, we use the explicit 
form of $C(t)$ for $q_{*}=1$ given in Ref.\cite{Mor:2015}. 
We expand $\xi(q_{*},p)$ and $c(q_{*},p)$ around
$p=p_{c}(q_{*})$ and take the limit $p\to p_{c}(q_{*})$ by
 fixing $x=t/\xi(q_{*},p)$. For the general $q_{*}>1/2$, 
we estimate $b(q_{*})$ using the expressions for $c(q_{*},p),\xi(q_{*},p)$ and 
the assumption in Eq.(\ref{eq:UC_D0}). For sufficiently large $x$, $g(x)\simeq 
\sqrt{4\pi x}$. As $\lim_{t\to \infty}C(t)=c(q,p)$, for 
$p\simeq p_{c}(q_{*})$, we obtain
\[
c(q_{*},p)=\sqrt{4\pi} b(q_{*})/\sqrt{\xi(q_{*},p)}.
\]
We can estimate $b(q_{*})$ by
\[
b(q_{*})=\lim_{p\to p_{c}(q_{*})}\frac{\sqrt{\xi(q_{*},p)}c(q_{*},p)}{\sqrt{4\pi}}.
\]

We have derived $g(x)$ exactly only for $q_{*}=1$. 
We are able to check $g(x)$ for $q_{*}<1$.
We can estimate $C(t)$ for $t\le 10^{5}$ and $q_{*}=0.6$ by numerically 
integrating the master equation for the system. 
Figure \ref{fig:UC_D} shows the results of this process. 
 The empty symbols indicate the results
for $p>p_{c}(q_{*})$. We have adopted the value of $p$ in the vicinity of 
$p_{c}(q_{*})=1/6$ and $t \in [10^{4},4\times 10^{5}]$.
As can be clearly seen, the data obtained for 
 different $p$ and $t$ values lies on the curve of Eq.(\ref{eq:UC_D}).
 The filled symbols indicate the results for $p<p_{c}(q_{*})$.
 The  curve of Eq.(\ref{eq:UC_D}) describe the data.
\end{document}